\documentclass[conference]{IEEEtran}
\usepackage{mathrsfs}
\usepackage{bbding}
\usepackage{graphicx}
\usepackage{float}
\usepackage{booktabs}
\usepackage{hyperref}
\usepackage{cite}

\usepackage{enumitem}
\usepackage{color}
\usepackage{ulem}
\usepackage{psfrag}
\usepackage{subfigure}
\usepackage{amssymb}
\usepackage{amsthm}
\usepackage{setspace}
\usepackage{epsfig}
\usepackage{pifont}
\usepackage{amsmath}
\usepackage{array}
\newcommand{\PreserveBackslash}[1]{\let\temp=\\#1\let\\=\temp}
\newcolumntype{C}[1]{>{\PreserveBackslash\centering}p{#1}}
\newcolumntype{R}[1]{>{\PreserveBackslash\raggedleft}p{#1}}
\newcolumntype{L}[1]{>{\PreserveBackslash\raggedright}p{#1}}
\usepackage{multicol}
\usepackage{multirow}
\usepackage{diagbox}
\usepackage{pifont}
\usepackage{indentfirst}
\usepackage{amsfonts}
\usepackage{algorithm}
\usepackage{algorithmicx}
\usepackage{algpseudocode}
\floatname{algorithm}{Procedure}
\usepackage{fancyhdr}
\usepackage{amscd}
\usepackage[cmyk]{xcolor}
\usepackage{bm}
\usepackage{fancyhdr}
\setlist{itemsep=0pt,parsep=0pt}

\newtheorem{proposition}{Proposition}

\newtheorem{remark}{Remark}

\allowdisplaybreaks[4]
\IEEEoverridecommandlockouts

\usepackage{caption}
\begin{document}
	\title{\huge Multi-User Continuous-Aperture Array Communications: \\How to Learn Current Distribution?\vspace{-4mm}}
	
	\author{
		%
		\IEEEauthorblockN{Jia Guo, Yuanwei Liu, and Arumugam Nallanathan}
		
		\IEEEauthorblockA{Queen Mary University of London\\ \{jia.guo, yuanwei.liu, a.nallanathan\}@qmul.ac.uk} \vspace{-9mm}}
	\maketitle
	\setcounter{page}{1}
	\thispagestyle{empty}
	
\begin{abstract}
	The continuous aperture array (CAPA) can provide higher degree-of-freedom and spatial resolution than the spatially discrete array (SDPA), where optimizing multi-user current distributions in CAPA systems is crucial but challenging. The challenge arises from solving non-convex functional optimization problems without closed-form objective functions and constraints.
 	In this paper, we propose a deep learning framework called L-CAPA to learn current distribution policies. In the framework, we find finite-dimensional representations of channel functions and current distributions, allowing them to be inputted into and outputted from a deep neural network (DNN) for learning the policy. 
	To address the issue that the integrals in the loss function without closed-form expressions hinder training the DNN in an unsupervised manner, we propose to design another two DNNs for learning the integrals. 
	The DNNs are designed as graph neural networks to incorporate with the permutation properties of the mappings to be learned, thereby improving learning performance. 
	Simulation results show that L-CAPA can achieve the performance upper-bound of optimizing precoding in the SDPA system as the number of antennas approaches infinity, and it is with low inference complexity.
	
	
%
\end{abstract}

\vspace{-1mm}
\section{Introduction}\label{sec:intro}
\vspace{-1mm}
The continuous aperture array (CAPA) communication is emerging as a paradigm shift from the conventional spatially discrete array (SDPA) and paves the way to next-generation antenna technologies \cite{zhao2024continuous}. By forming with a continuous electromagnetic aperture, the CAPA can be regarded as a SPDA comprising an infinite number of antennas with infinitesimal spacing, thereby providing high spatial degree-of-freedom \cite{liu2023near}. 

The current distribution on the CAPA is the key factor affecting the system performance \cite{Comm_LIS_2020}, which can be seen as precoding for SDPA systems. The optimization of current distribution is challenging, because functional optimization problems need to be solved to find the optimal distribution, where traditional convex optimization tools are not applicable.
Existing works found optimal solutions to the problem under single- and two-user cases \cite{zhao2024continuous}, which can hardly be extended to systems with more users. In \cite{zhang2023pattern}, the challenge of functional optimization was addressed by leveraging the Fourier basis of functions, such that the original problem can be converted to a parameter optimization problem that is easier to solve. A sub-optimal solution to the functional optimization problem can be found by dividing the CAPA into multiple patches such that the problem can be reduced to one in SDPA system, but the discretization yields significant performance loss \cite{sanguinetti2022wavenumber}. 

Noticing that deep neural networks (DNNs) are powerful for learning unknown relationships from massive training data, and have been successfully applied to learn wireless policies such as precoding \cite{GNN_BF_MISO_TWC_2024} and power allocation \cite{GJ_TWC_GNN} in SPDA systems, we strive to learn the current distribution policy (i.e., the relationship from channels to current distribution) with DNNs. Nonetheless, directly applying DNNs to learn the policy encounters two issues. Firstly, both the channels and current distributions are functions that can be regarded as vectors with infinite dimensions, which cannot be directly inputted into or outputted from DNNs. Secondly, supervised learning cannot be adopted for training the DNNs because the solution of the optimization problem is not available to serve as labels. If the DNNs are trained with unsupervised learning by letting the objective function of the optimization problem be the training loss, then the integrals in the objective function without closed-form expression hinder back-propagation.

In this paper, we propose a deep learning framework called L-CAPA concerning these two issues for learning current distributions. To resolve the first issue, we find finite-dimensional representation vectors of the channel functions and current distributions that can be inputted into and outputted from a DNN. By regarding the second issue, we are motivated by the model-free deep learning method proposed in \cite{SCJ-ModelFree}. Specifically,  another two DNNs are trained to learn the integrals in the objective functions and constraints, through which the gradients can be back-propagated to learn the current distribution policy. The permutation properties of the mappings to be learned are further incorporated with the DNNs for reducing the cost of training and improving learning performance \cite{GJ_TWC_GNN}. As far as the authors know, this is the first paper concerning learning current distribution in CAPA systems. 
Simulation results show that the proposed framework for learning a spectral efficiency (SE)-maximal current distribution can achieve the performance upper bound of optimizing precoding in the SDPA system when the number of antennas approaches infinity. When the number of antennas is not large, the proposed framework achieves much higher SE than the optimization-based method with much shorter inference time, which enables real-time implementation.

\vspace{-1mm}
\section{\!System Model and Current Distribution Policy}\label{sec: system model}
\vspace{-1mm}
Consider a downlink system where a base station (BS) transmits to $K$ users, with both the BS and each user equipped with a CAPA. The aperture spaces of the CAPAs at the BS and the $k$-th user are denoted as $\mathcal{A}$ and $\mathcal{A}_k\subseteq {\mathbb R}^{3\times 1}$, respectively, and the CAPA of the $k$-th user is centered at $\mathbf{s}_k \in {\mathbb R}^{3\times 1}$.

\vspace{-1mm}
\subsection{System Model}
The transmitted signal at point $\mathbf{r}\in\mathcal{A}$ can be expressed as $x(\mathbf{r})=\sum_{k=1}^K \mathsf{V}_k(\mathbf{r})s_k$, where $s_k$ is the symbol to be transmitted to the $k$-th user with ${\mathbb E}\{|s_{k}|^2\}=1$, and $\mathsf{V}_k(\mathbf{r}) (\mathbf{r}\in\mathcal{A})$ is the continuous current distribution to convey $s_k$. When the CAPA is discretized to SDPA, the current distribution degenerates to the precoding vector of each user. 

The signal-to-interference-and-noise ratio if the $k$-th user can be approximated as \cite{zhao2024continuous},
\begin{eqnarray}\label{eq:sinr}
	\gamma_k 
	\approx
	\frac{|\mathcal{A}_k|\cdot|\int_{\mathcal{A}} \mathsf{H}_k(\mathbf{r}) \mathsf{V}_k(\mathbf{r}) d\mathbf{r}|^2}{\sum_{j=1,j\neq k}^K |\mathcal{A}_j|\cdot|\int_{\mathcal{A}} \mathsf{H}_k(\mathbf{r}) \mathsf{V}_j(\mathbf{r}) d\mathbf{r}|^2 + \sigma_k^2},
\end{eqnarray}
where $\mathsf{H}_k(\mathbf{r})$ is the channel response of the center of the $k$-th user's aperture from point $\mathbf{r}$ of the BS's aperture, $\sigma_k^2$ satisfies ${\mathbb E}\{\mathsf{N}_k(\mathbf{s})\mathsf{N}_k^*(\mathbf{s}')\}=\sigma_k^2\delta(\mathbf{s}-\mathbf{s}')$, $\mathsf{N}_k(\mathbf{s})$ is the thermal noise at the point $\mathbf{s}$ of the $k$-th user's aperture, $(\cdot)^*$ denotes conjugate of a complex value, $|\mathcal{A}_k|$ is the aperture size of the $k$-th user, and the approximation comes from the assumption that the aperture size of each user is smaller enough compared to the propagation distance and the BS aperture size. 

When there are only line-of-sight (LoS) channels between the BS and the users, $\mathsf{H}_k(\mathbf{r})$ can be modeled as follows \cite{zhao2024continuous},
\begin{eqnarray}\label{eq:chn-model}
		\mathsf{H}_k(\mathbf{r}) \!\!&\!\!=\!\!&\!\! \sqrt{\frac{\mathbf{e}_r^T(\mathbf{s}_k-\mathbf{r})}{\|\mathbf{r}-\mathbf{s}_k\|}}\cdot\notag\\
		&&\frac{jk_0\eta e ^{-jk_0\|\mathbf{r\!-\!s}_k\|}}{4\pi\|\mathbf{r}-\mathbf{s}_k\|}\Bigg(\!1\!+\!\frac{j/k_0}{\|\mathbf{r}-\mathbf{s}_k\|}\!-\!\frac{1/k_0^2}{\|\mathbf{r}-\mathbf{s}_k\|^2}\!\Bigg)\!,
\end{eqnarray}
where  ${\mathbf{e}_r}\in{\mathbb{R}}^{3\times1}$ is the normal vector of the CAPA at the BS, $\eta=120\pi$ is the impedance of free space, $k_0=\frac{2\pi}{\lambda}$ with $\lambda$ being the wavelength denotes the wavenumber, and $(\cdot)^T$ denotes matrix transpose.

\vspace{-1mm}
\subsection{Current Distribution Optimization Problem and Policy}

With \eqref{eq:sinr}, we can formulate a problem to optimize the current distributions for all the users. Taking maximizing SE as an example, the problem can be formulated as,
\begin{subequations}
	\begin{align}
		{\bf P1}:
		\max_{\mathsf{V}_k(\mathbf{r}),k=1,\cdots,K}~~ & \textstyle\sum_{k=1}^K \log_2 \left(1+\gamma_k \right) \label{eq:bb-object} \\
		{\mathrm{s.t.}} ~~
		&  \textstyle\sum_{k=1}^K\int_{\mathcal{A}}|\mathsf{V}_k(\mathbf{r})|^2 d\mathbf{r} =  P_{\max},\label{eq:bb-constraint}
	\end{align}
\end{subequations}
where $P_{\max}$ is the maximal transmit power of the BS.

The current distribution policy is referred to as the mapping from the space channel response to the current distribution, which is denoted as a function $\mathsf{V}(\cdot)=F_{\mathsf{CD}}(\mathsf{H}(\cdot))$, where $\mathsf{V}(\cdot)=[\mathsf{V}_1(\cdot),\cdots,\mathsf{V}_K(\cdot)]$ and $\mathsf{H}(\cdot)=[\mathsf{H}_1(\cdot),\cdots,\mathsf{H}_K(\cdot)]$ are multi-variate functions. We do not express the current distribution policy as $\mathsf{V}(\mathbf{r})=F_{\mathsf{CD}}(\mathsf{H}(\mathbf{r}))$, which may cause misunderstanding that the current distribution at point $\mathbf{r}$ is only related to the channel response at this point.

\vspace{-1mm}
\section{Learning to Optimize Current Distribution}
\vspace{-1mm}

In this section, we propose a deep learning-based method to learn the current distribution policy $\mathsf{V}(\cdot)=F_{\mathsf{CD}}(\mathsf{H}(\cdot))$. We first show the issues encountered when directly applying conventional deep learning framework as in \cite{GJ_TWC_GNN}, and then illustrate how to resolve these issues with our proposed framework.

\begin{figure*}[!htb]
	\centering
	\includegraphics[width=\linewidth]{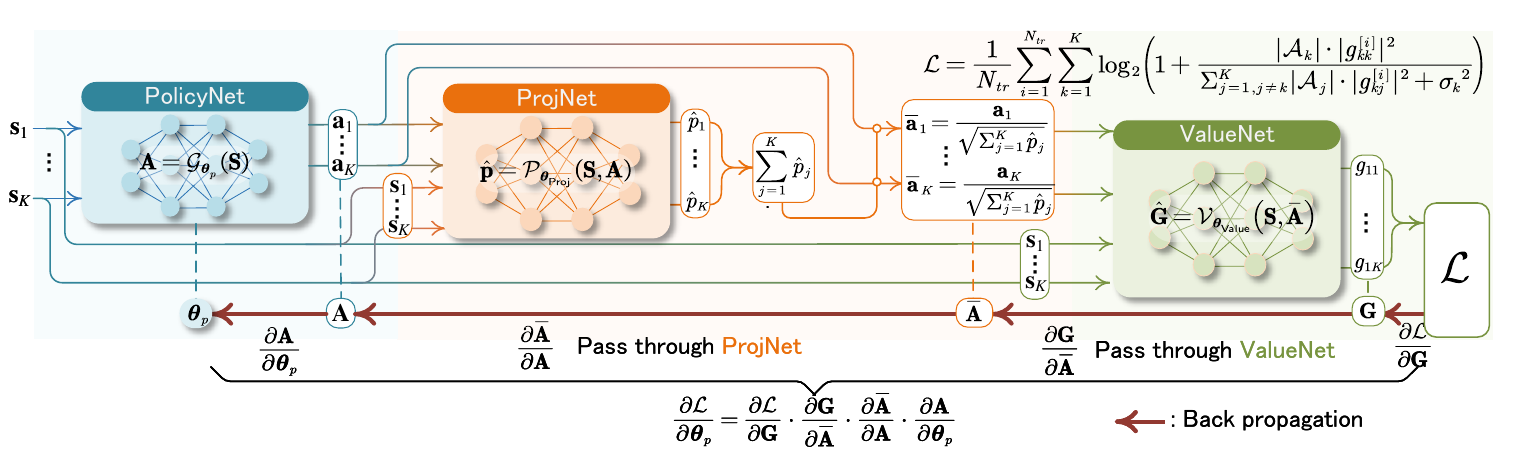}
	\vspace{-5mm}
	\caption{Illustration of L-CAPA.}
	\vspace{-6mm}
	\label{fig:dl-arch}
\end{figure*}

\vspace{-1mm}
\subsection{\!Issues of Applying Conventional Deep Learning Framework}
Intuitively, a DNN can be applied to learn the current distribution policy, where the input and output of the DNN are respectively  $\mathsf{H}(\cdot)$ and $\mathsf{V}(\cdot)$. Since the learned current distribution may not satisfy the power constraint in \eqref{eq:bb-constraint}, we normalize the learned distribution as follows such that it can be projected to the region satisfying the constraint,
\begin{equation}\label{eq:act-func}
	\bar{\mathsf{V}}_k(\mathbf{r})=\mathsf{V}_k(\mathbf{r})\sqrt{P_{\max}}\Big/\sqrt{\textstyle\sum_{j=1}^K \textstyle\int_{\mathcal{A}}|\mathsf{V}_j(\mathbf{r})|^2 d\mathbf{r}}.
\end{equation}

The DNN cannot be trained in a supervised manner as in \cite{L2O}, because the optimal (or even near-optimal) solution of problem \textbf{P1} is not available to serve as the label for training. If the DNN is instead trained in an unsupervised manner as in \cite{GJ_TWC_GNN}, the loss function should be the objective function in \eqref{eq:bb-object} averaged over all the training samples, i.e., 

\vspace{-2mm}
{\small
\begin{equation}\label{eq:loss-func}
	\mathcal{L}\!=\!\frac{1}{N_{\mathsf{tr}}}\!\sum_{i=1}^{N_{\mathsf{tr}}}\!\sum_{k=1}^K\!\log_2\!\Bigg(\!1+\frac{|\mathcal{A}_k|\cdot|\int_{\mathcal{A}} \mathsf{H}_k^{[i]}(\mathbf{r}) \bar{\mathsf{V}}_k^{[i]}(\mathbf{r}) d\mathbf{r}|^2}{\sum\limits_{j=1,j\neq k}^K |\mathcal{A}_j|\!\cdot\!|\int_{\mathcal{A}} \mathsf{H}_k^{[i]}(\mathbf{r}) \bar{\mathsf{V}}_j^{[i]}(\mathbf{r}) d\mathbf{r}|^2 \!+\! \sigma_k^2}\Bigg),
\end{equation}}

\noindent where $N_{\mathsf{tr}}$ denotes the number of training samples, and the superscript $[i]$ denotes the $i$-th sample.

However, such a design encounters two issues as follows.
\begin{itemize}[leftmargin=0.4cm]
	\item \emph{Issue 1: $\mathsf{H}(\cdot)$ and $\mathsf{V}(\cdot)$ are functions that can be regarded as vectors with infinite dimensions, whereas DNNs can only receive inputs and give outputs with finite dimensions}. 
	\item \emph{Issue 2: Both \eqref{eq:act-func} and \eqref{eq:loss-func} include integrals that are without closed-form expressions}. This hinders computing the gradients (specifically, $\frac{\partial \mathcal{L}}{\partial \bar{\mathsf{V}}^{[i]}_k(\cdot)}$ and $\frac{\partial\bar{\mathsf{V}}^{[i]}_k(\cdot)}{\partial \mathsf{V}^{[i]}_k(\cdot)}$) for training the DNN (i.e., back-propagation).
\end{itemize}

\vspace{-1mm}
\subsection{The Proposed Deep Learning Framework}\label{sec:proposed-framework}
To resolve these two issues, we propose a deep learning framework to learn current distribution policies, which is called learning to communicate for CAPA (L-CAPA). The core idea for resolving Issue 1 is to find finite-dimensional representations of the functions that can be inputted into and outputted from DNNs. The core idea for resolving Issue 2 is to learn the integrals with DNNs. Since the input-output relationships of DNNs only include basic operations such as matrix addition and multiplication, the relationships are with closed-form expression, such that the gradients can be computed for learning the policy.

The overall proposed framework is shown in Fig. \ref{fig:dl-arch}, which contains (i) a PolicyNet to learn a finite-dimensional representation of $\mathsf{V}(\cdot)$ from a finite-dimensional representation of $\mathsf{H}(\cdot)$, (ii) a ProjNet to learn the integral in \eqref{eq:act-func}, and (iii) a ValueNet to learn the integrals in \eqref{eq:loss-func}. Both the ProjNet and ValueNet are designed such that the gradients of the loss function can be computed to train the PolicyNet. In the following, we will introduce the neural networks in detail.

\subsubsection{PolicyNet}\label{sec:policynet}
The PolicyNet is used to learn the mapping from $\mathsf{H}(\cdot)$ to $\mathsf{V}(\cdot)$. To this end, we need to resolve \emph{Issue 1} such that only finite-dimensional vectors are inputted into and outputted from the PolicyNet. One way to resolve the problem is to find the ``bases'' of these two functions.

\textbf{Finding finite-dimensional representation of $\mathsf{H}(\cdot)$:} It can be seen from the channel model \eqref{eq:chn-model} that $\mathsf{H}_k(\mathbf{r})$ only depends on  $\mathbf{e}_r, \eta, k_0$ and the locations of the $k$-th user's aperture, i.e., $\mathbf{s}_k$. Since $\mathbf{e}_r, \eta, k_0$ can be seen as constants in the CAPA transmission system, $\mathsf{H}_k(\mathbf{r})$ only varies with $\mathbf{s}_k$, and $\mathsf{H}_k(\mathbf{r})$ is known if $\mathbf{s}_k$ is known. Hence, $\mathbf{s}_k\in{\mathbb R}^{3\times 1}$ can be used as a finite-dimensional representation of $\mathsf{H}_k(\mathbf{r})$.

\textbf{Finding finite-dimensional representation of $\mathsf{V}(\cdot)$:} To this end, we provide the following proposition.
\begin{proposition}\label{prop1}
	The optimal current distribution is in a function subspace spanned by $\mathsf{H}_1(\mathbf{r}),\cdots,\mathsf{H}_K(\mathbf{r})$, i.e., $\mathsf{V}_k(\mathbf{r})$ can be expressed as $\mathsf{V}_k(\mathbf{r})=\sum_{j=1}^K a_{jk} \mathsf{H}_j(\mathbf{r})$.
	\begin{IEEEproof}
		See Appendix \ref{proof:prop1}.
	\end{IEEEproof}
\end{proposition}

This proposition indicates that once $\mathsf{H}_1(\mathbf{r}),\cdots,\mathsf{H}_K(\mathbf{r})$ are known, we can obtain $\mathsf{V}_k(\mathbf{r})$ with $\mathbf{a}_k\triangleq [a_{1k},\cdots,a_{Nk}]^T$, which is called the \emph{weighting vector} of the $k$-th user. Hence, $\mathbf{a}_k$ can be used as a finite-dimensional representation of $\mathsf{V}_k(\mathbf{r})$.

After finding the finite-dimensional representations of  $\mathsf{H}(\cdot)$ and $\mathsf{V}(\cdot)$, we can learn the current distribution policy by learning $\mathbf{A}\triangleq[\mathbf{a}_1,\cdots,\mathbf{a}_K]\in {\mathbb C}^{K\times K}$ from $\mathbf{S}\triangleq[\mathbf{s}_1,\cdots,\mathbf{s}_K]^T\in {\mathbb C}^{K\times 3}$. Then, the policy to be learned is denoted as $\mathbf{A}^\star=F_{\mathsf{P}}(\mathbf{S})$, where $\mathbf{A}^\star$ is the finite-dimensional representation that corresponds to the optimal current distribution. 

When the order of users is changed by a permutation matrix $\bm\Pi$, $\mathbf{S}$ is permuted to ${\bm\Pi}^T\mathbf{S}$. We can see from problem \textbf{P1} that both the objective function and the constraint in the problem keep unchanged. Then, only the order of the optimal current distributions of the $K$ users changes while the values remain unchanged. This indicates that $\mathbf{A}^\star$ is permuted to ${\bm\Pi}^T\mathbf{A}^\star{\bm\Pi}$ after permutation of users. Hence, the current distribution policy satisfies the following permutation property,
\begin{eqnarray}\label{eq:perm-policy}
	{\bm\Pi}^T\mathbf{A}{\bm\Pi} = F_{\mathsf{P}}({\bm\Pi}^T\mathbf{S}).
\end{eqnarray}

\textbf{Neural Network Structure:} The input and output of the PolicyNet are respectively $\mathbf{S}$ and $\mathbf{A}$. The PolicyNet can be a fully-connected neural network (FNN), or a graph neural network (GNN) that leverages the permutation property in \eqref{eq:perm-policy} for low training complexity and potential of generalizability to different problem scales (e.g., number of users), as to be detailed in section \ref{sec:gnn}. The input-output relationship of the PolicyNet is denoted as,
\begin{equation}\label{eq:PolicyNet}
	\mathbf{A}=\mathcal{G}_{\bm\theta_p}(\mathbf{S}),
\end{equation}  
where $\bm\theta_p$ denotes all the free parameters (e.g., the weight matrices) in the PolicyNet.

The training of $\bm\theta_p$ depends on the trained ProjNet and ValueNet. Hence, we first introduce these two neural networks, and then introduce how to train the PolicyNet in section \ref{sec:training}.

\subsubsection{ProjNet}
The learned finite-dimensional representations of current distribution (i.e., the weighting vectors) need to be normalized as in \eqref{eq:act-func} such that it is projected to the feasible region of problem \textbf{P1}. To resolve Issue 2 that the integral in \eqref{eq:act-func} (i.e., the power consumed by $\mathsf{V}_k(\mathbf{r})$) hinders back-propagation, we design the ProjNet to learn power consumption as follows. 

The ProjNet aims to learn the mapping from the finite-dimensional representations of $\mathsf{V}_k(\mathbf{r}),k=1,\cdots,K$ to the power consumption for each user. As have been analyzed in section \ref{sec:policynet}, $\mathsf{H}_k(\mathbf{r})$ is known if $\mathbf{s}_k$ is known, and after $\mathsf{H}_k(\mathbf{r}),k=1,\cdots,K$ are known, we can obtain $\mathsf{V}_k(\mathbf{r}),k=1,\cdots,K$ with $\mathbf{A}$. Then, $\mathbf{S}$ and $\mathbf{A}$ can be the representation of $\mathsf{V}_k(\mathbf{r}),k=1,\cdots,K$, and the mapping to be learned by the ProjNet is denoted as $\mathbf{p}=F_{\mathsf{Proj}}(\mathbf{S,A})$, where $\mathbf{p}=[p_1,\cdots,p_K]^T$ is the vector of power consumption for the $K$ users. With the same way of analyzing the permutation property of the current distribution policy in \eqref{eq:perm-policy}, we can obtain the property satisfied by the function $F_{\mathsf{Proj}}(\cdot)$ as,
\begin{equation}\label{eq:perm-proj}
	{\bm\Pi}^T\mathbf{p}=F_{\mathsf{Proj}}({\bm\Pi}^T\mathbf{S}, {\bm\Pi}^T\mathbf{A}{\bm\Pi}).
\end{equation}

\textbf{Input and output}: To learn the mapping $\mathbf{p}=F_{\mathsf{Proj}}(\mathbf{S,A})$, the inputs are $\mathbf{S}$ and $\mathbf{A}$, and the output is $\hat{\mathbf{p}}\triangleq[\hat{p}_1,\cdots,\hat{p}_K]^T$, where $\hat{p}_k$ is the learned power consumption by $\mathsf{V}_k(\mathbf{r})$.

\textbf{Neural network structure}: The ProjNet can also be designed as a GNN, as to be detailed in section \ref{sec:gnn}. The input-output relationship of the PolicyNet is denoted as,
\begin{equation}\label{eq:ProjNet}
	\hat{\mathbf{p}}=\mathcal{P}_{\bm\theta_{\sf Proj}}(\mathbf{S,A}),
\end{equation}
where $\bm\theta_{\sf Proj}$ denotes all the free parameters  in the ProjNet.

\textbf{Training of the ProjNet}: The ProjNet can be trained with $N_{\mathsf{tr}}$ samples in a supervised manner. The $i$-th training sample is generated as $\{\mathbf{S}^{[i]}, \mathbf{A}^{[i]}, \mathbf{p}^{[i]}\}$, where $\mathbf{p}^{[i]}=[p_1^{[i]},\cdots, p_k^{[i]}]^T$ is the expected output of power consumption in the $i$-th sample. Since $p_k^{[i]}$ is without closed-form expression, it can be computed numerically by discretizing the integral region $\mathcal{A}$ into multiple sub-regions and approximate the integral with summation, i.e.,
\begin{equation}\label{eq:cal-integral}
	p_k^{[i]} = \textstyle\int_{\mathcal{A}} |\mathsf{V}_k^{[i]}(\mathbf{r})|^2 d\mathbf{r} \approx \textstyle\sum_{m=1}^M |\mathsf{V}_k^{[i]}(\mathbf{r}_m)|^2 \Delta,
\end{equation}
where $\mathbf{r}_m$ is the center of the $m$-th sub-region, and $\Delta$ is the area of each sub-region. $\mathsf{V}_k^{[i]}(\mathbf{r})$ is the current distribution of the $k$-th user in the $i$-th sample, which is obtained by $\mathsf{V}_k^{[i]}(\mathbf{r})=\sum_{j=1}^K a_{jk}^{[i]} \mathsf{H}_k^{[i]}(\mathbf{r})$, and $\mathsf{H}_k^{[i]}(\mathbf{r})$ is obtained by substituting $\mathbf{s}_k^{[i]}$ into the channel model in \eqref{eq:chn-model}.

After generating the training samples, the neural network is trained with the following loss function to minimize the difference between the actual outputs and the expected outputs,
\begin{equation}\label{eq:loss-proj}
	\mathcal{L}({\bm\theta}_{\mathsf{Proj}}) = \textstyle\frac{1}{N_{\mathsf{tr}}}\sum_{i=1}^{N_{\mathsf{tr}}}\sum_{k=1}^K \big(\hat{p}_k^{[i]} - {p}_k^{[i]}\big)^2.
\end{equation}

\begin{remark}
	Although the integral in \eqref{eq:cal-integral} is approximated with summation by discretizing the integral region, the learned current distribution is a continuous function, which is different from the optimization-based methods that only optimize precoding on discontinuous points in the region (e.g., the method ``MIMO'' for comparison in \cite{sanguinetti2022wavenumber}). 
\end{remark}

After the ProjNet is trained, given the input of $\mathbf{S}$ and $\mathbf{A}$, the ProjNet can output $\hat{\mathbf{p}}$. Then, $\mathbf{a}_k$ can be projected as follows to satisfy the power constraint \eqref{eq:bb-constraint},
\begin{equation}\label{eq:proj}
	\bar{\mathbf{a}}_k = \mathbf{a}_k\sqrt{P_{\max}}\Big/\sqrt{\textstyle\sum_{j=1}^K \hat{p}_j}. 
\end{equation}


\subsubsection{ValueNet}
To resolve Issue 2 that the integrals in \eqref{eq:loss-func} without closed-form expressions hinder training the PolicyNet in an unsupervised manner,
 we design a DNN to learn the integrals in \eqref{eq:bb-object}. The DNN is called ValueNet, because the outputs of the DNN are used to compute the value (i.e., loss function) with the learned policy.

The ValueNet aims to learn the mapping from the finite dimensional representations of $\mathsf{H}_k(\mathbf{r})$ and $\mathsf{V}_k(\mathbf{r}), k=1,\cdots,K$ to the integrals $g_{kj}\triangleq\int_{\mathcal{A}}\mathsf{H}_k(\mathbf{r})\mathsf{V}_j(\mathbf{r})d\mathbf{r}, k,j=1,\cdots,K$,
which is denoted as $\mathbf{G}=F_{\mathsf{Value}}(\mathbf{S},\bar{\mathbf{A}})$, where $\bar{\mathbf{A}}=[\bar{\mathbf{a}}_1,\cdots,\bar{\mathbf{a}}_K]$, $\mathbf{G}\in {\mathbb C}^{K\times K}$ with the element in the $k$-th row and the $j$-th column being $g_{kj}$. With the same way of analyzing the permutation property in \eqref{eq:perm-policy}, we can obtain the property satisfied by $F_{\mathsf{Value}}(\cdot)$ as,
\begin{equation}\label{eq:perm-value}
	{\bm\Pi}^T\mathbf{G}{\bm\Pi}=F_{\mathsf{Value}}({\bm\Pi}^T\mathbf{S},{\bm\Pi}^T\bar{\mathbf{A}}{\bm\Pi}).
\end{equation}

\textbf{Input and output}: To learn the function $\mathbf{G}=F_{\mathsf{Value}}(\mathbf{S},\bar{\mathbf{A}})$, the inputs of the ValueNet are $\mathbf{S}$ and $\bar{\mathbf{A}}$, and the output is $\hat{\mathbf{G}}\triangleq[\hat{g}_{kj}]\in{\mathbb C}^{K\times K}$, where $\hat{g}_{kj}$ is the approximated $g_{kj}$.

\textbf{Neural Network Structure}: The neural network can also be designed as a GNN, as to be detailed in section \ref{sec:gnn}. The input-output relationship of the ValueNet is denoted as,
\begin{equation}\label{eq:ValueNet}
	\hat{\mathbf{G}}=\mathcal{V}_{\bm\theta_{\sf Value}}(\mathbf{S,\bar{A}}).
\end{equation}

\textbf{Training of the ValueNet}: Similar to the ProjNet, the ValueNet can be trained with $N_{\sf tr}$ samples in a supervised manner. 
The $i$-th training sample is denoted as $\{\mathbf{S}^{[i]}, \bar{\mathbf{A}}^{[i]}, \mathbf{G}^{[i]}\}$, where the element in the $k$-th row and $j$-th column of $\mathbf{G}^{[i]}$, $g_{kj}^{[i]} = \int_{\mathcal{A}} \mathsf{H}_k^{[i]}(\mathbf{r}){\mathsf{V}}_j^{[i]}(\mathbf{r}) d\mathbf{r}$, can be generated by discretizing the integral region $\mathcal{A}$ into multiple sub-regions and approximate integral with summation, as how $p_k^{[i]}$  is generated in \eqref{eq:cal-integral}.

After generating the training samples, the ValueNet is trained with a loss function to minimize the difference between the actual outputs and the expected outputs, which is similar to the loss function in \eqref{eq:loss-proj} and hence is not provided here due to limited space.



\vspace{-1mm}
\subsection{Training and Inference Procedure} \label{sec:training}

The training procedure of the proposed framework is as follows. The ProjNet and ValueNet are firstly trained with supervised learning to obtain ${\bm \theta}_{\mathsf{Proj}}^\star$ and ${\bm \theta}_{\mathsf{Value}}^\star$. After the two neural networks are trained, 
given the input of $\mathbf{S}$ and $\bar{\mathbf{A}}$, the matrix of approximated integrals can be obtained as,
\begin{equation}\label{eq:output}
	\hspace{-2.1mm}\hat{\mathbf{G}}\!=\!\mathcal{V}_{\bm\theta_{\sf Value}^\star}(\mathbf{S,\bar{A}}) 
	\!\overset{(a)}{=}\! \mathcal{V}_{\bm\theta_{\sf Value}^\star}\!\Bigg(\!\mathbf{S},\!\frac{\mathcal{G}_{\bm\theta_p}(\mathbf{S})\sqrt{P_{\max}}}{\sqrt{\!\sum_{j\!=\!1}^K \!\mathcal{P}_{\bm\theta_{\sf Proj}^{\star}}\!\big(\mathbf{S},\mathcal{G}_{\bm\theta_p}(\mathbf{S}) \big)_{\! j}}}\!\Bigg)\!,\!
\end{equation}
where $\bm\theta_{\sf Proj}^{\star}$ and $\bm\theta_{\sf Value}^{\star}$ are respectively the trained free parameters of the ProjNet and the ValueNet, $(a)$ comes from substituting \eqref{eq:proj}, \eqref{eq:ProjNet} and \eqref{eq:PolicyNet} into the equation, $\mathcal{P}_{\bm\theta_{\sf Proj}^{\star}}\big(\cdot)_j$ denotes the $j$-th element in the output of the function. We can see that $\hat{\mathbf{G}}$ only depends on $\bm\theta_p$ after the ProjNet and ValueNet are trained.
Then, the loss function for training the PolicyNet can be computed as follows, which is the objective function averaged over all the training samples,
\begin{equation}
	\mathcal{L}(\bm\theta_p)=\frac{1}{N_{\mathsf{tr}}}\sum_{i=1}^{N_{\mathsf{tr}}} \sum_{k=1}^K\log_2\Bigg(1+\frac{|\mathcal{A}_k|\cdot|\hat{g}_{kk}^{[i]}|^2}{\sum_{j=1,j\neq k}^K |\mathcal{A}_j||\hat{g}_{kj}^{[i]}|^2 + \sigma_k^2}\Bigg). \notag
\end{equation}
With the loss function, the gradients can be computed to update $\bm\theta_p$ with a step size of $\phi$ as follows,
{\small
\begin{equation}
	\bm\theta_{p} \leftarrow \bm\theta_{p}-\phi\frac{\partial \mathcal{L}(\bm\theta_{p})}{\partial \bm\theta_{p}} = \bm\theta_{p} - \frac{\phi}{N_{\sf tr}}\sum_{i=1}^{N_{\sf tr}}\frac{\partial \mathcal{L}(\bm\theta_{p})}{\partial \mathbf{G}^{[i]}}\cdot \frac{\partial \mathbf{G}^{[i]}}{\partial \bar{\mathbf{A}}^{[i]}}\cdot \frac{\partial \bar{\mathbf{A}}^{[i]}}{\partial \mathbf{A}^{[i]}}\cdot \frac{\partial \mathbf{A}^{[i]}}{\partial \bm\theta_p}, \notag\\
\end{equation} }

\vspace{-2mm}
\noindent where $\frac{\partial \mathbf{G}^{[i]}}{\partial \bar{\mathbf{A}}^{[i]}}$ and $\frac{\partial \bar{\mathbf{A}}^{[i]}}{\partial \mathbf{A}^{[i]}}$ are respectively passed through the trained ValueNet and ProjNet, as shown in Fig. \ref{fig:dl-arch}.

In the inference phase, given the input of $\mathbf{S}$, the matrix of weighting vectors $\mathbf{A}$ can be outputted by the trained PolicyNet. Then, $\mathbf{A}$ can be projected to $\bar{\mathbf{A}}$ that satisfies \eqref{eq:bb-constraint} with the trained ProjNet, and the learned current distribution can be obtained as $\bar{\mathsf{V}}_k(\mathbf{r})=\sum_{j=1}^K \bar{a}_{jk} \mathsf{H}_k(\mathbf{r})$. The ValueNet is no longer required in this phase, because it is only used to obtain the integrals in the loss function for training the PolicyNet.

\vspace{-1mm}
\section{\!Learning Current Distribution over Graph} \label{sec:gnn}
\vspace{-1mm}
In L-CAPA, three DNNs need to be trained to learn three mappings that respectively satisfy the permutation properties in \eqref{eq:perm-policy}, \eqref{eq:perm-proj} and \eqref{eq:perm-value}. In this section, we model the multi-user CAPA system as a graph and learn the three mappings  with GNNs, such that the permutation properties can be satisfied.

\begin{figure}[!htb]
	\centering
	\includegraphics[width=\linewidth]{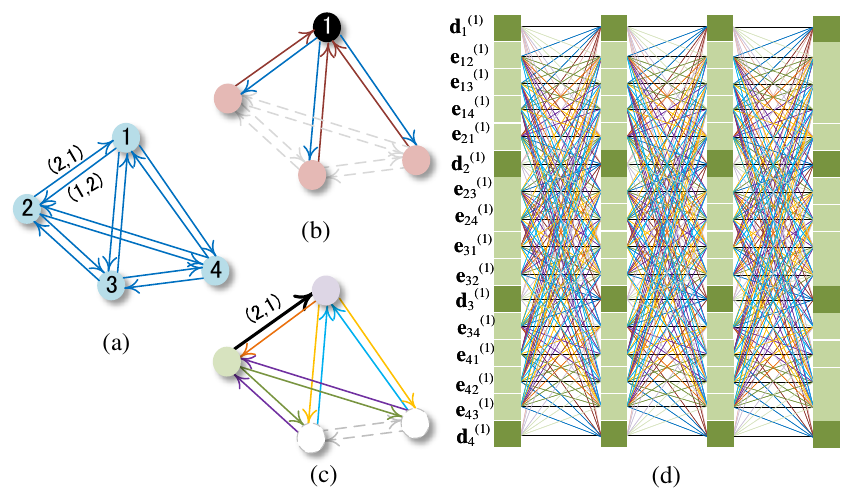}
	\vspace{-7mm}
	\caption{Illustration of (a) a directed-graph with four vertices, (b) updating the representation of the first vertex, (c) updating the representation of edge $(2,1)$, and (d) a GNN with four layers. In (b), (c) and (d), the same color indicates the same weight matrix, and the dashed circle and arrow mean that the representation of the vertex or edge is not aggregated.}
	\vspace{-2mm}
	\label{fig:fig-gnn}
\end{figure}

The multi-user CAPA system can be modeled as a directed graph as shown in Fig. \ref{fig:fig-gnn}(a), where each user is a vertex. There are directed edges between every two vertices. Denote the edge from the $k$-th vertex to the $j$-th vertex as edge $(k,j)$, where the two vertices are called the neighboring vertices of the edge $(k,j)$, and the edge $(k,j)$ is called the neighboring edge of the $k$-th vertex and the $j$-th vertex. 

There are features and actions on each vertex or edge, which are listed below for the three mappings to be learned.
\begin{itemize}
	\item \textbf{Mapping of PolicyNet (i.e., $\mathbf{A}^\star=F_{\mathsf{P}}(\mathbf{S})$)}:
	\begin{itemize}
		\item \emph{Features}: The feature of the $k$-th vertex is $\mathbf{s}_k$, and there are no features on the edges.
		\item \emph{Actions}: The action of the $k$-th vertex is $a_{kk}$, and the action of edge $(k,j)$ is $a_{kj}$.
	\end{itemize}
	\item \textbf{Mapping of ProjNet (i.e., $\mathbf{p}=F_{\mathsf{Proj}}(\mathbf{S,A})$)}:
	\begin{itemize}
		\item \emph{Features}: The feature of the $k$-th vertex is $\{\mathbf{s}_k, a_{kk}\}$, and the feature of edge $(k,j)$ is $a_{kj}$.
		\item \emph{Actions}: The action of the $k$-th vertex is $\hat{p}_k$, and there are no actions on the edges.
	\end{itemize}  
	\item \textbf{Mapping of ValueNet (i.e., $\mathbf{G}=F_{\mathsf{Value}}(\mathbf{S},\bar{\mathbf{A}})$)}:
	\begin{itemize}
		\item \emph{Features}: The feature of the $k$-th vertex is $\{\mathbf{s}_k, \bar{a}_{kk}\}$, and the feature of edge $(k,j)$ is $\bar{a}_{kj}$.
		\item \emph{Actions}: The action of the $k$-th vertex is $\hat{g}_{kk}$, and the action of edge $(k,j)$ is $\hat{g}_{kj}$.
	\end{itemize}  
\end{itemize}

GNNs are used to learn the mappings from features to actions. Although the features and actions for learning the three mappings are different, we can use three GNNs with the same structure to learn the three mappings, because they are learned over the same graph. 

Since at least one of the features and the actions is defined on both the vertices and edges, we propose a GNN with $L$ layers, where both the representations of the vertices and edges are updated in each layer with an \emph{update equation}. To update the representation of each vertex or each edge $(\ell+1)$-th layer, the representations of neighboring edges and vertices in the previous layer are firstly processed by a \emph{processor} to extract information from the representations. Then, the extracted information is aggregated with a \emph{pooling function}, and combined with the representation of the vertex or edge itself with a \emph{combiner}. When the processor is linear, the pooling function is summation, and the combiner is a linear function cascaded by an activation function $\sigma(\cdot)$, the update equation can be written as (taking updating the representation of the $k$-th vertex as an example),
\begin{equation}
	\mathbf{d}_k^{(\ell+1)} \!=\! \sigma\Big(\mathbf{W}_k\mathbf{d}_k^{(\ell)} \!+\! \textstyle\sum_{i=1,i\neq k}^K\!(\mathbf{W}_i\mathbf{d}_i^{(\ell)} + \mathbf{W}_{ik}\mathbf{e}_{ik}^{(\ell)}+ \mathbf{W}_{ki}\mathbf{e}_{ki}^{(\ell)})\Big), \notag
\end{equation} 
where $\mathbf{W}_*$ is the weight matrix, $\mathbf{d}_k^{(\ell)}$ and $\mathbf{e}_{kj}^{(\ell)}$ are respectively the representation of the $k$-th vertex and edge $(k,j)$. In the input layer (i.e., $\ell\!=\!1$) and output layer (i.e., $\!\ell\!=\! L$), $\mathbf{d}_k^{(\ell)}$($\mathbf{e}_{kj}^{(\ell)}$) are respectively the features and actions of the vertices (edges).

Parameter sharing can be introduced into the weight matrices of the GNNs to satisfy the permutation properties. The update equation and the parameter sharing scheme are omitted here due to limited space, and an example of parameter sharing is provided in Fig. \ref{fig:fig-gnn}(b) and (c). 
Fig. \ref{fig:fig-gnn} shows the structure of a GNN with four layers.

\vspace{-1mm}
\section{Simulation Results}
\vspace{-1mm}

Consider that a BS deployed with a planar CAPA transmits to $K$ users. The aperture size of the CAPA is $|\mathcal{A}|=4~m^2$ unless otherwise specified. Each user's position can be expressed with a spherical coordinate $(r,\theta,\phi)$. The users are deployed in a region where $20m<r<30m$, $\frac{\pi}{6}<\theta<\frac{\pi}{3}$ and $\frac{\pi}{6}<\phi<\frac{\pi}{3}$. The normal vector of the CAPA and the wavelength are respectively set as $\mathbf{e}_r=[0,1,0]$ and $\lambda=0.0107$ m \cite{zhao2024continuous}. Without loss of generality, we set $|\mathcal{A}_k|=\frac{\lambda^2}{4\pi}$ and $\sigma_k=\sigma_0,\forall k$. We control the signal-to-noise  ratio (SNR) of the system by changing $\zeta\triangleq\frac{|\mathcal{A}_k|}{\sigma_0^2}\frac{k_0^2\eta^2}{4\pi}$.

To show the impact of incorporating with the permutation property on the learning performance, we compare the performances of two methods, where the PolicyNet, the ProjNet and the ValueNet are respectively set as GNNs and FNNs without incorporating with any properties, and the two methods are referred to as ``L-CAPA (GNN)'' and ``L-CAPA (FNN)''.


We compare the performance of the learning-based methods with a numerical algorithm. Since problem \textbf{P1} is hardly to be solved because there are integrals included in the objective functions and constraints, we discretize the CAPA into $M$ squared regions, and the channel response in each region is approximated by the channel response at the center of each region. In this case, the original problem is reduced to a precoding optimization problem in the SDPA system, where the precoding vectors for the users can be optimized with the WMMSE algorithm \cite{WMMSE2011Shi}. The optimal precoding vector is also a weighted sum of channel vectors. By multiplying the weighted vectors with the $\mathsf{H}_1(\mathbf{r}),\cdots, \mathsf{H}_K(\mathbf{r})$, we can obtain an approximated solution of problem \textbf{P1}. The solution serves as the performance baseline of the learning-based methods.

\begin{figure}[!htb]
	\centering
	\begin{minipage}[t]{0.45\linewidth}
		\centering
		\includegraphics[width=\textwidth]{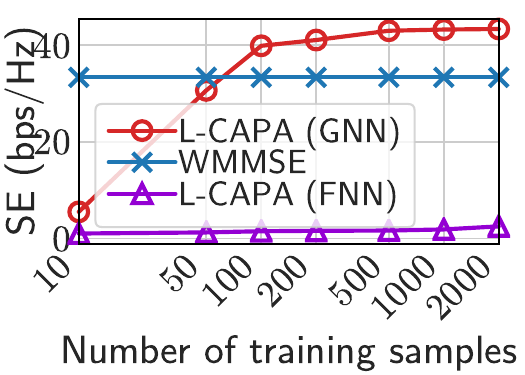}
		\vspace{-6mm}
		\caption{SE versus $N_{\mathsf{tr}}$, $K=4, M=256$, SNR=60 dB.}
		\label{fig:perf-ntr}
	\end{minipage}\hspace{3mm}
	\begin{minipage}[t]{0.45\linewidth}
		\centering
		\includegraphics[width=\textwidth]{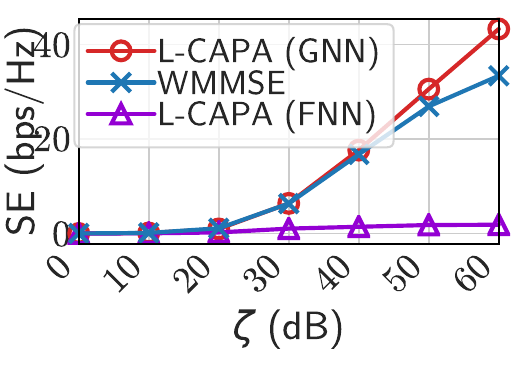}
		\vspace{-6mm}
		\caption{SE versus SNR, $K=4, M=256$, $N_{\mathsf{tr}}=2000$. }
		\label{fig:perf-snr}
	\end{minipage}
	
	\begin{minipage}[t]{0.45\linewidth}
		\centering
		\includegraphics[width=\textwidth]{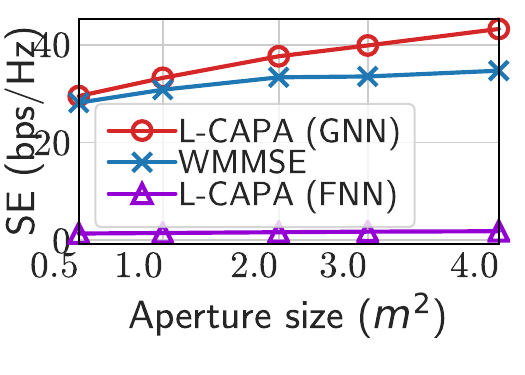}
		\vspace{-6mm}
		\caption{SE versus aperture size, $K=4, M=256$, SNR=60 dB. }
		\label{fig:perf-apsize}
	\end{minipage}\hspace{3mm}
	\begin{minipage}[t]{0.45\linewidth}
		\centering
		\includegraphics[width=\textwidth]{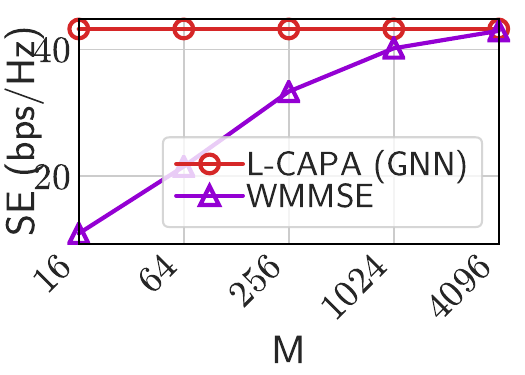}
		\vspace{-6mm}
		\caption{SE versus $M$, $K=4$, SNR=60 dB. }
		\label{fig:perf-wm}
	\end{minipage}
	\vspace{-3mm}
\end{figure}

Fig. \ref{fig:perf-ntr} $\sim$ Fig. \ref{fig:perf-apsize} respectively show the achieved SE under different numbers of training samples, values of SNR and aperture sizes, where the the number of discretized squared regions (i.e., $M$) for the WMMSE algorithm is not large for affordable computational complexity. We can see that the SE achieved by ``L-CAPA (GNN)'' is much higher than the SE achieved by ``L-CAPA (FNN)''. The performance gain comes from leveraging the permutation properties. The results in Fig. \ref{fig:perf-ntr} also indicate that the number of samples required for ``L-CAPA (GNN)'' to achieve an expected performance (i.e., sample complexity) is much lower than ``L-CAPA (FNN)''.
We can also see from the figures that ``L-CAPA (GNN)'' performs better than the WMMSE algorithm when SNR is large. This is because the solution after discretization is only a sub-optimal solution of problem \textbf{P1}.

In Fig. \ref{fig:perf-wm}, we show how the SE of the WMMSE algorithm changes with $M$. For comparison, we also provide the performance of ``L-CAPA (GNN)'' with enough training samples (according to Fig. \ref{fig:perf-ntr}, 2000 samples are enough for a good learning performance). It can be seen that the SE grows slower with $M$ and is close to but does not exceed the SE achieved by ``L-CAPA (GNN)'', which indicates that the learned current distribution can achieve the performance upper bound of optimizing precoding in the SDPA system when $M\to\infty$. 

We next compare the average inference time of ``L-CAPA (GNN)'' with the average running time of WMMSE algorithm over all the test samples. When $K=4, |\mathcal{A}|=4$ m$^2$, the inference time of ``L-CAPA (GNN)'' is 0.02 s, while the running time of the WMMSE algorithm is 9.6 s, which is much longer due to computing the integrals in \eqref{eq:act-func} for satisfying the power constraint.


\vspace{-2mm}
\section{Conclusions}
\vspace{-1mm}
In this paper, we proposed a deep learning framework called L-CAPA to learn current distribution policies in CAPA systems. In the framework, a ProjNet and a ValueNet were first trained to learn the integrals in the constraint and the objective function of the optimization problem. Then, the trained DNNs help back-propagating gradients to a PolicyNet that learns the current distribution policy. The three neural networks were designed as GNNs to leverage the permutation properties of the mappings to be learned. Simulation results demonstrated the performance of L-CAPA in terms of achieving high SE with short running time. The proposed framework can also be applied to learn other policies in the CAPA system, such as power allocation and user scheduling, and the designed GNNs have the potential of generalizing to different problem scales (e.g., number of users), which are left for future works.
%
%
\vspace{-2mm}
\begin{appendices}
	\numberwithin{equation}{section}
	\vspace{-1mm}
	\section{Proof of Proposition \ref{prop1}}\label{proof:prop1}\vspace{-1mm}
		We prove this with proving its converse-negative proposition: if $\exists k$ such that the optimal $\mathsf{V}_k(\mathbf{r})$ cannot be expressed as $\mathsf{V}_k(\mathbf{r})=\sum_{j=1}^K a_{jk} \mathsf{H}_j(\mathbf{r})$, then there exists a better solution that can achieve higher SE while satisfying  \eqref{eq:bb-constraint}.
	
	If $\exists k$ such that the optimal  $\mathsf{V}_k(\mathbf{r})$ is not in the subspace spanned by $\mathsf{H}_1(\mathbf{r}),\cdots,\mathsf{H}_K(\mathbf{r})$ (called channel subspace for short in the sequel), then $\mathsf{V}_k(\mathbf{r})$ can be expressed as
	\begin{equation}\label{eq:perp}
		\mathsf{V}_k(\mathbf{r})=\mathsf{V}_{k//}(\mathbf{r})+\mathsf{V}_{k\perp}(\mathbf{r}),
	\end{equation}
	where $\mathsf{V}_{k//}(\mathbf{r})$ is in the channel subspace, while $\mathsf{V}_{k\perp}(\mathbf{r})$ does not, i.e., $\int_{\mathcal{A}}\mathsf{V}_{k\perp}(\mathbf{r})\mathsf{H}_j(\mathbf{r})d\mathbf{r}=0 , \forall j$, and
	\begin{equation}\label{eq:perp1}
		\textstyle\int_{\mathcal{A}}\mathsf{V}_{k\perp}(\mathbf{r})\mathsf{V}_{k//}(\mathbf{r})d\mathbf{r}=0.
	\end{equation}
	In this case, the SE of all the users can be expressed as,
	\begin{equation}
		R_0=\sum_{j=1}^K \log_2\Bigg(1+\frac{|\mathcal{A}_j|\!\cdot\!|\int_{\mathcal{A}} \mathsf{H}_j(\mathbf{r}) \mathsf{V}_j'(\mathbf{r}) d\mathbf{r}|^2}{\sum_{i=1,i\neq j}^K\! |\mathcal{A}_i|\!\cdot\!|\int_{\mathcal{A}} \mathsf{H}_j(\mathbf{r}) \mathsf{V}_i'(\mathbf{r}) d\mathbf{r}|^2 \!+\! \sigma_i^2}\Bigg),\notag
	\end{equation}
	where 
	\begin{equation}\label{eq:perp2}
		\mathsf{V}_j'(\mathbf{r})=\left\{
		\begin{aligned}
			\mathsf{V}_j(\mathbf{r}), & ~~j \neq k, \\
			\mathsf{V}_{j//}(\mathbf{r}), &~~ j=k.
		\end{aligned}
		\right.
	\end{equation}
	
	With \eqref{eq:perp} and \eqref{eq:perp1}, we have $\int_{\mathcal{A}}|\mathsf{V}_k(\mathbf{r})|^2d\mathbf{r}=\int_{\mathcal{A}}|\mathsf{V}_{k//}|^2(\mathbf{r})d\mathbf{r}+\int_{\mathcal{A}}\mathsf|{V}_{k\perp}(\mathbf{r})|^2d\mathbf{r}+2\int_{\mathcal{A}}|\mathsf{V}_{k//}(\mathbf{r})\mathsf{V}_{k\perp}(\mathbf{r})|d\mathbf{r}=\int_{\mathcal{A}}|\mathsf{V}_{k//}(\mathbf{r})|^2 d\mathbf{r}+\int_{\mathcal{A}}|\mathsf{V}_{k\perp}(\mathbf{r})|^2 d\mathbf{r}$, hence $\int_{\mathcal{A}}|\mathsf{V}_k(\mathbf{r})|^2d\mathbf{r}>\int_{\mathcal{A}}|\mathsf{V}_{k//}(\mathbf{r})|^2 d\mathbf{r}$. 
	
	We now construct a solution of \textbf{P1} as
	$\bar{\mathsf{V}}_j(\mathbf{r}) = C\mathsf{V}_j'(\mathbf{r})$ with\vspace{-3mm}
	\begin{equation}
		C = \sqrt{\frac{P_{\max}}{\sum_{j=1}^K \int_{\mathcal{A}}|\mathsf{V}_j'(\mathbf{r})|^2 d\mathbf{r}}} = \sqrt{\frac{\sum_{j=1}^K \int_{\mathcal{A}}|\mathsf{V}_j(\mathbf{r})|^2 d\mathbf{r}}{\sum_{j=1}^K \int_{\mathcal{A}}|\mathsf{V}_j'(\mathbf{r})|^2 d\mathbf{r}}}.\notag
	\end{equation}
	This is a feasible solution of \textbf{P1}, because it satisfies constraint \eqref{eq:bb-constraint}. From \eqref{eq:perp2} and $\int_{\mathcal{A}}|\mathsf{V}_k(\mathbf{r})|^2d\mathbf{r}>\int_{\mathcal{A}}|\mathsf{V}_{k//}(\mathbf{r})|^2 d\mathbf{r}$, we can obtain that $C>1$. Then, the SE achieved by the constructed solution is
	{\small
		\begin{eqnarray}
			R_1\!\!\!&\!\!\!=\!\!\!&\!\!\!\sum_{j=1}^K \log_2\Bigg(1+\frac{|\mathcal{A}_j|\cdot|\int_{\mathcal{A}} \mathsf{H}_j(\mathbf{r}) \mathsf{V}_j'(\mathbf{r})\cdot C d\mathbf{r}|^2 }{\sum_{i=1,i\neq j}^K |\mathcal{A}_i|\cdot|\int_{\mathcal{A}} \mathsf{H}_j(\mathbf{r}) \mathsf{V}_i'(\mathbf{r})\cdot C d\mathbf{r}|^2 + \sigma_i^2}\Bigg) \notag\\
			\!\!\!&\!\!\!=\!\!\!&\!\!\!\sum_{j=1}^K\! \log_2\!\Bigg(\!1\!+\!\frac{|\mathcal{A}_j|\cdot|\int_{\mathcal{A}} \mathsf{H}_j(\mathbf{r}) \mathsf{V}_j'(\mathbf{r}) d\mathbf{r}|^2 }{\sum_{i=1,i\neq j}^K |\mathcal{A}_i|\!\cdot\!|\int_{\mathcal{A}} \mathsf{H}_j(\mathbf{r}) \mathsf{V}_i'(\mathbf{r}) d\mathbf{r}|^2 \!+\! \frac{\sigma_i^2}{C^2}}\!\Bigg)\!> \! R_0. \notag
		\end{eqnarray}
	}
	
	\noindent This indicates that the feasible solution we constructed can achieve a higher SE than $\mathsf{V}_1(\mathbf{r}),\cdots,\mathsf{V}_K(\mathbf{r})$, hence $\{\mathsf{V}_1(\mathbf{r}),\cdots,\mathsf{V}_K(\mathbf{r})\}$ is not the optimal solution.
	
\end{appendices}

\vspace{-2mm}
\bibliography{IEEEabrv,GJ}
\vspace{-2mm}

\end{document}